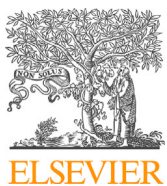
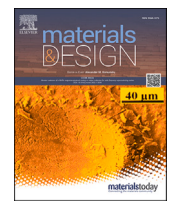

# Mo-Re-W alloys for high temperature applications: Phase stability, elasticity, and thermal property insights via multi-cell Monte Carlo and machine learning

Tyler D. Doležal [a,b,*], Nick A. Valverde [c], Jodie A. Yuwono [d], Ryan A. Kemnitz [e]

[a] *Department of Materials Science and Engineering, Massachusetts Institute of Technology, Cambridge, MA, USA*
[b] *Department of Engineering Physics, Air Force Institute of Technology, Wright-Patterson Air Force Base, OH, USA*
[c] *Department of Physics and Astronomy, Michigan State University, East Lansing, MI, USA*
[d] *School of Chemical Engineering, The University of Adelaide, Adelaide, SA 5005, Australia*
[e] *Department of Aeronautics and Astronautics, Air Force Institute of Technology, Wright-Patterson Air Force Base, OH, USA*

## ARTICLE INFO

Dataset link: https://github.com/SaminGroup/Dolezal-MC2

Dataset link: https://github.com/tylerdolezal

## ABSTRACT

The increasing demand for materials capable of withstanding high temperatures and harsh environments necessitates the discovery of advanced alloys. This study introduces a computational routine to predict solid-state phase stability and calculates elastic constants to determine high temperature viability. With it, machine learning models were trained on 1,014 Mo-Re-W structures to enable a large compilation of elastic and thermal properties over the complete Mo-Re-W compositional domain with extreme resolution. A series of heat maps spanning the full compositional domain were generated to visually present the impact of alloy constituents on the alloy properties. Our findings identified a balanced (Mo,W) + Re blend as a promising composition for high temperature applications, attributed to a strong and stable (Mo,W) matrix with high Re content and the formation of strengthening (W,Re) precipitates that enhanced mechanical performance at 1600 °C. Several Mo-Re-W compositions were manufactured to experimentally validate the computational predictions. This approach provides an efficient and system-agnostic pathway for designing and optimizing alloys for high-temperature applications.

## 1. Introduction

As a community of material engineers and scientists we stand on the precipice of an era of rapid discovery with an almost boundless number of new materials waiting to be explored. Near-future industry and government mechanical systems demand higher operating temperatures and harsher operating conditions, generating a strong need for advanced structural solutions. The material research community is actively responding to these demands for high-temperature systems. Examples include hypersonic propulsion engines [1], next-generation nuclear reactors [2,3], kinetic defense systems [4,5], nuclear fusion reactors [6,7], and nuclear-powered space propulsion systems [8].

Hypersonic flight propulsion systems face extreme temperatures near 1800 °C and beyond while simultaneously exposed to an oxidizing atmosphere [1]. A front-running candidate for hypersonic systems are high entropy alloys (HEAs) [9] containing refractory metals. Comprehensive data on refractory HEAs was compiled by Couzinié et al. [10], with on-going progress reported in recent studies [11–14]. These alloys exhibit excellent high-temperature mechanical properties, high melting points [14], and exceptional oxidation resistance due to complex protective oxide layers [15–18].

Next-generation nuclear power plants will operate at higher temperatures and use highly corrosive molten salts [2,3]. Structural materials must resist corrosion, oxidation, and irradiation [3]. ORNL introduced a Ni-based Hastelloy for its Molten Salt Reactor Experiment [2], still considered a promising candidate despite issues with He bubble formation and alloy swelling [19]. Recent research into oxide dispersion-strengthened (ODS) nickel-based alloys shows promise for overcoming these challenges [20]. Transition metal borides (TMB$_2$) are also under extensive study as additives to enhance mechanical properties of

**Table 1**
(a) In-text reference name and respective metallic composition, by atomic percent (at%), of the trial alloys discussed throughout this work. (b) The temperature in-text reference name along with their values in Celsius and Kelvin.

| (a) | at% | | | (b) | Temperature | |
|---|---|---|---|---|---|---|
| Reference | Mo | Re | W | Reference | °C | K |
| a | 82 | 6 | 12 | LT | 26.85 | 300 |
| b | 66 | 13 | 21 | HT | 1600 | 1873.15 |
| c | 45 | 16 | 39 | | | |
| d | 33 | 8 | 59 | | | |

structural materials for high temperature applications, especially in Ni superalloys. [21,22]

Tungsten and tungsten heavy alloys (WHAs) are crucial for defense systems [4,5], fusion targets [23], fusion reactor designs [6,7], and nuclear-powered space propulsion systems [8] due to their superior thermal conductivity and high temperature strength. Although tungsten is brittle and difficult to alloy, recent advancements in manufacturing techniques offer new possibilities [24]. The addition of rhenium (Re) to tungsten increases ductility and mechanical strength while reducing brittleness [25–30]. Furthermore, molecular dynamics studies have shown promising results for HEAs in extreme environments [31].

As demonstrated by Lai, Kishore, Yu, and their collaborators [29–31], computational tools can theoretically probe alloy performance in extreme environments. In recent years, several computational methods have been developed to study new materials. Niu et al. presented a multi-cell Monte Carlo routine for predicting alloy crystal structures [32,33]. Osei-Agyemang and Balasubramanian introduced a methodology for studying early-stage oxidation mechanisms in HEAs using first principles DFT [34–36]. Wei et al. explored the magnetic behavior of FeCoNiSi$_{0.2}$Cr$_{0.2}$ and FeCoNiSi$_{0.2}$Mn$_{0.2}$ HEAs [37]. Li et al. employed DFT calculations to characterize the impact of adding transition metals (TMs), such as Mo, W, and Re, to brittle materials and concluded the addition of TMs enhanced mechanical performance, reduced hardness, and promoted ductility. [38] Liu et al., developed a magneto-mechanical model based on density functional theory to assess the relationship between weak magnetic signals and critical damage in pipelines due to stress, analyzing the effects of different stress directions and materials on these signals and verifying the findings experimentally [39].

Here, we present a cooperative effort between experimental and computational communities using our derivative [40] of the multi-cell Monte Carlo method [32], referred to as (MC)$^2$, to predict phase stability and characterize a set of Mo-Re-W alloys as possible candidates for extreme high temperature aerospace applications [41]. The procedure enabled the development of machine learning models which provided DFT-free elastic constants predictions. The purpose of this work is to demonstrate and verify a system-agnostic and application-agnostic computational procedure that helps guide the physical manufacturing of promising high temperature materials, chosen from a vast compositional domain, to satisfy a multitude of high-temperature applications.

## 2. Methodology

A graphical representation of the procedure is provided in Fig. 1. A set of four alloys containing Mo, Re, and W was explored. The alloys were examined at room temperature (26.85 °C), referred to as LT for low temperature, and 1600 °C, referred to as HT for high temperature. The reference names, compositions, and temperatures have been provided in Table 1. The solid-state phase of a-d was determined at LT and HT via (MC)$^2$. Elastic properties of the phases were calculated via the optimized high efficiency strain-matrix sets (OHESS) procedure [42] as implemented in ElasTool [43]. These theoretical predictions were compared against experimental measurements made on real alloy samples.

### 2.1. Predicting the alloy phase

(MC)$^2$ was executed using three super simulation cells initialized in the conventional crystal structures of Mo, W, and Re, which were body centered cubic (BCC), BCC, and hexagonal close packed (HCP), respectively. Altogether, eight (MC)$^2$ simulations were performed (four alloys at two different temperatures). Each simulation cell contained 48 atoms for a total of 144 atoms throughout the entire simulation and an atomic percent (at%) fidelity of 1/48. One Monte Carlo step began with attempting an atomic species flip. This move is described as randomly selecting one of the three simulation cells, then randomly selecting one atom within the chosen simulation cell and "flipping" it from its current species type to one of the other two species types. The acceptance criterion, based on the Metropolis criterion [44], for the flip is given in Eq (1).

$$P_{accept}^{flip} = min\left\{1, \exp\left(-\beta\left[\Delta H(1-w) + w\frac{\delta_{ij}}{2}\Delta \tilde{H}\right] + N\Delta G\right)\right\}, \quad (1)$$

where $\Delta H$, $\Delta \tilde{H}$, and $\Delta G$ are defined as,

$$\Delta H = m\sum_{i=1}^{m}(U'_i + pV'_i)f'_i - m\sum_{i=1}^{m}(U_i + pV_i)f_i \quad (2)$$

$$\Delta \tilde{H} = m\sum_{i=1}^{m}(U'_i - U_i) \quad (3)$$

$$\Delta G = \sum_{i=1}^{m}[f'_i \ln(V'_i) - f_i \ln(V_i)]$$
$$+ \sum_{i=1}^{m} f'^i \sum_{j=1}^{m} X'^i_j \ln(X'^i_j) - \sum_{i=1}^{m} f^i \sum_{j=1}^{m} X^i_j \ln(X^i_j). \quad (4)$$

Here, $\beta = 1/k_BT$, where $k_B$ is the Boltzmann constant, N is the sum of all the particles across all simulation cells, $m$ is the total number of simulation cells, $U_i$ is the energy of simulation cell $i$, $V_i$ is the volume of simulation cell $i$, $p$ is the pressure, which was set to 0, and $f_i$ is the molar fraction of simulation cell $i$. Letting $n^i_j$ be the number of species $i$ in simulation cell $j$, the atomic concentration is given by $X^i_j = n^i_j / \sum_{k=1}^{3} n^k_j$, which is the atomic concentration of species $i$ in simulation cell $j$. The primed coordinates indicate post-flipped values. The term $\delta_{ij}$ is the number of flips performed to change an atom of type $i$ to type $j$. For example, if two atoms of type A were changed into type B, $\delta_{AB} = 2$. For this work, the number of flips was always 1:1, and so $\delta_{ij} = 1$. The term $w$ is a parameter that is tuned to inflict an energetic penalty cost on the acceptance of a new state. It was determined that, for $w$ values less than 0.60, the enthalpy of mixing contribution ($X \ln X$) to the overall acceptance factor would overrule the fact that the total system energy would increase. In this way, the simulation is essentially "dragged" away from the lowest energy state in favor of an increased molar fraction value for the simulation cell with the lowest energy. The upper bound for $w$ is in good agreement with a foundational (MC)$^2$ paper [33]. The molar fraction of the new phases was calculated using the Lever rule to enforce mass conservation.

Spin-polarized DFT calculations were executed to obtain the electronic energy and volume changes of the simulation cell. The calculations were performed using the Projector Augmented Wave (PAW) method as implemented by VASP. The calculations were performed with a plane wave cutoff energy of 400 eV and a 3x3x2 Monkhorst-Pack [45] k-point mesh. Spin-polarized DFT calculations performed on the simulation cells allowed for changes in the volume and atomic positions through the setting ISIF = 3. The electronic self-consistent calculation was converged to 1x10$^{-6}$ eV and ionic relaxation steps were performed using the conjugate-gradient method (IBRION = 2) and continued until the total force on each atom dropped below a tolerance of 1x10$^{-2}$ eV/Å. The generalized gradient approximation (GGA) was used for the exchange correlation functionals as parameterized by Perdew-Burke and Ernzerof (PBE) [46]. The PAW pseudopotentials [36] were used with the valence electron configurations $4p^65s^14d^5$, $6s^25d^4$, and $5p^66s^25d^5$ for Mo, W, and Re, respectively. Please note, unless otherwise stated, DFT calculations were executed following these settings.

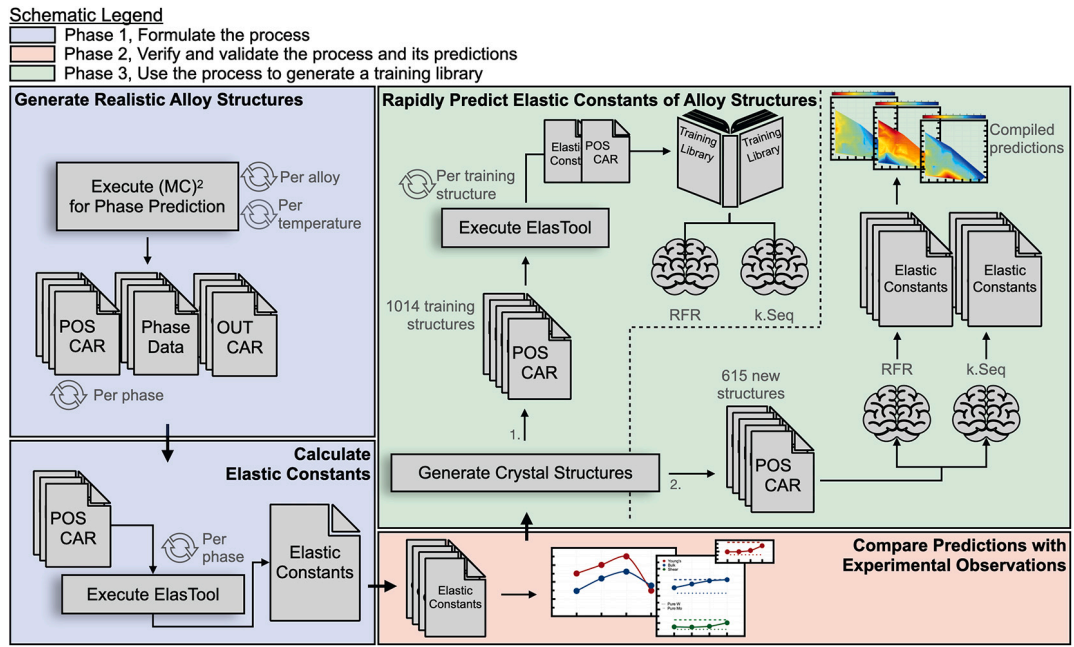

**Fig. 1.** A graphical representation of the prediction and characterization routine. The schematic legend explains the color-coordination of the three main stages: Phase 1 (blue) involved predicting realistic phases using (MC)$^2$ and calculating the elastic properties with ElasTool. Phase 2 (orange) compared these predictions with experimental observations to ensure validity. Phase 3 (green) focused on rapidly predicting elastic constants using ElasTool, training machine learning models, and generating predictions. Dashed lines in Phase 3 separate the steps for building the training library and generating 615 elastic constant predictions, resulting in elastic moduli heat maps. Either the random forest regressor (RFR) or sequential deep learning (k.Seq) model can be used to generate the final heat maps.

### 2.2. Elastic and thermal properties

The elastic properties were computed at LT and HT via the optimized high efficiency strain-matrix sets (OHESS) procedure [42] as implemented in ElasTool [43,47]. Deformations to the crystal lattice were introduced via ElasTool's automated routine using four strains for the cell deformations: -0.06, -0.03, 0.03, and 0.06. For HT elastic predictions, VASP ab initio NpT MD simulations were executed using a 2x2x2 Monkhorst-Pack k-point mesh for 500 time steps with a time step width of 2 femtoseconds.

Under the Voight-Reuss-Hill approximation the bulk (B), shear (G), and Young's (Y) moduli, and Pugh ratio (G/B) were calculated and compared. The reader is referenced to Ref. [43,48,47] for an open access detailed description of elastic property relationships. The micro-hardness of an alloy, H, was calculated using the empirical expression [49]

$$H = 0.096 \left( \frac{1 - 8.5\nu + 19.5\nu^2}{1 - 7.5\nu + 12.2\nu^2 + 19.6\nu^3} \right) \times Y, \quad (5)$$

where $\nu$ is the Poisson ratio. The micro-hardness offers insights on material strength and microstructural characteristics, ensuring quality control, and determining the suitability of materials for specific applications, particularly those involving high loads or abrasive environments. The elastic response and atomic bond behavior of each alloy was considered using the Cauchy pressure (C") [50] and the Kleinman parameter ($\xi$) [51]. Cauchy pressure is defined in Eq. (6),

$$C'' = C_{12} - C_{44}, \quad (6)$$

where a positive value indicates metallic bonding, defined by the interaction of metallic atoms with the electron gas of their nearest neighbors. A negative C" value indicates covalent bonding characteristics and a zero C" value indicates a system governed by a central force potential [52].

Furthermore, a positive C" value indicates a more ductile material and one which is more tolerant to damage [53]. Pettifor's rule [54] indicates that positive C" leads to increased ductility due to metallic bonding while negative C" leads to a brittle material due to covalent bonding. The Kleinman parameter was calculated according to Eq. (7),

$$\xi = \frac{C_{11} + 8C_{12}}{7C_{11} + 2C_{12}}, \quad (7)$$

which ranges from 0 to 1. The parameter describes bonding behavior between bonds bending and bonds stretching. A material with $\xi$ closer to 1 tends to deform more easily through changes in the bond angles, i.e. bending, while a material with $\xi$ closer to 0 is more easily deformed through bond length changes, i.e. stretching. Together, C" and $\xi$ were examined against the Pugh ratio (G/B) to see if a general relationship could be deduced between a material's ductility and these calculated values.

To complement the analysis of a material's ductility, the local-lattice distortions (LLD) approach for classifying a multi-principal-element alloy's brittle or ductile behavior was considered here [55]. The LLD is defined in Eq. (8),

$$\text{LLD} = \Delta w_{VEC} \times \frac{\Delta u_{x,y,z}}{||u_{x,y,z}||} \quad (8)$$

$$\begin{cases} \text{LLD} < 0.3 & \text{ductile} \\ \text{LLD} \geq 0.3 & \text{brittle} \end{cases},$$

where $\Delta w_{VEC} = VEC_j^{bcc} - (VEC_{max}^{bcc} - VEC_{min}^{bcc})$ with $VEC_{min}^{bcc} = 4$ and $VEC_{max}^{bcc} = 6$. The valence electron count for phase $j$ is given by $VEC_j^{bcc} = \sum_{i=1}^{3} x_j^i VEC^i$, where $VEC^i$ is the valence electron count of alloy constituent $i$. For example, a phase with 70% Mo and 30% Re would have a $VEC_{phase}^{bcc} = (0.70 \times 6) + (0.30 \times 7) = 6.3$. The atomic displacement is defined as the difference in an atom's lattice position relative to the perfect BCC lattice position it occupied. The average displacement over all the atoms is $\Delta u_{x,y,z}$ and the $L_2$ norm of the displacement vectors is $||u_{x,y,z}||$. To avoid finite-cell displacement effects the LLD structures were 5x5x5 BCC Super-Cell Random APproximates (SCRAPs) at the concentrations predicted by (MC)$^2$. VASP relaxation calculations were executed as previously described with the exception that a 3x3x3 Monkhorst-Pack k-point mesh was used. The same SCRAPs were used to analyze the stacking fault energy (SFE) of the alloy systems by introducing a [110] stacking fault to the upper $\hat{z}$ region of the simulation cell. The affected atoms were held fixed while the rest of the cell was relaxed. The SFE was calculated according to Eq. (9), where $E_{fault}$ and $E_0$ were the energies of the faulted structure and unperturbed structure, respectively, and A was the area of the stacking fault plane,

$$\text{SFE} = \frac{E_{fault} - E_0}{A} \,. \tag{9}$$

Analyzing the SFE provides insights into the mechanical properties of the alloy. SFE affects the ease of dislocation movement, which in turn influences the material's ductility, strength, and work-hardening behavior. Lower SFE typically leads to increased ductility and work hardening, while higher SFE can result in stronger, but more brittle materials. Understanding the SFE helps in optimizing the alloy composition to achieve desired mechanical properties for high-temperature and high-stress applications.

The Debye Temperature ($\Theta_D$) was calculated according to Eq. (10) and has dependencies on the mass of the material constituents, the crystal structure, and the electronic interactions along the lattice.

$$\Theta_D = v_m \frac{h}{k_B} \left[\frac{3n}{4\pi}\left(\frac{N_A \rho}{M}\right)\right]^{1/3} \tag{10}$$

In Eq. (10), $v_m$ is average wave velocity, defined in Eq. (11), $h$ is Planck's constant, $k_B$ is Boltzmann's constant, $N_A$ is Avogadro's number, n in the number of atoms in the simulation cell, M is the mass of the simulation cell, and $\rho$ is the cell density.

$$v_m = \left[\frac{1}{3}\left(\frac{1}{v_l^3} + \frac{2}{v_s^3}\right)\right]^{-1/3} \tag{11}$$

The average wave velocity depends on the longitudinal ($v_l$) and shear ($v_s$) wave velocities which are defined in Eq. (12) and Eq. (13), respectively.

$$v_l = \sqrt{\frac{B + 4G/3}{\rho}} \tag{12}$$

$$v_s = \sqrt{\frac{G}{\rho}} \tag{13}$$

The values for $\Theta_D$ can be used to link thermal properties with elastic properties, making it a valuable parameter to analyze when considering high temperature applications. A high $\Theta_D$ generally indicates strong interatomic forces and a stiff lattice, which typically leads to higher thermal conductivity due to reduced phonon scattering. An increase in $\Theta_D$ can also suggest stronger bonds between atoms, as higher energy phonons correspond to stiffer bonds. To complement this thermal analysis, the minimum thermal conductivity ($k_{min}$), as given by the Cahill and Clarke model [56,48], was also considered. The $k_{min}$ is a value that is approached when a material exists at temperatures above $\Theta_D$ and is unique in the fact that it is independent of defects due to the reduced mean free path of phonons at such high temperatures. Finally, the melting temperature was predicted using the empirical relation given in Eq. (14) [57],

$$T_m = 354\text{K} + 4.5\frac{\text{K}}{\text{GPa}}\left(\frac{2C_{11} + C_{33}}{3}\right) \pm 350 \text{ K} \,. \tag{14}$$

In this way, generating thermal property predictions was possible without needing to execute costly DFT calculations to generate the dynamical matrix of a structure.

### 2.3. Machine learning

To circumvent the need to prepare and run slow and costly DFT calculations for elastic constant and thermal property predictions, a machine-learning-based approach was developed. Essentially, the model acts as an "elastic constants calculator", delivering elastic constants in seconds. From these, a large number of elastic and thermal properties can be calculated and compiled for a holistic analysis on each constituent's contribution to the alloy properties. This approach enabled the rapid scanning of the complete Mo-Re-W compositional domain with a fidelity of 2 at.%, including many structural configurations at each composition for statistically meaningful elastic considerations.

To generate a training set of symmetrically nonequivalent structures, the derivative enumeration algorithm [58] was used. A total of 1,014 48-atom BCC structures were used with varied amounts of each species to account for abundant and dilute mixtures of each constituent. The elastic constants of each structure was calculated using ElasTool. An independent verification set of 33 48-atom BCC derivative structures at a wide range of Mo-Re-W compositions was held back from the training set to examine each models general accuracy and performance. The elastic constants of the verification set were calculated in the same way as the training set.

The first machine learning algorithm was implemented using the Keras API provided through Python. The keras.Sequential() model (k.Seq) was employed with a linear stack of five connected layers. The number of neurons in each hidden layer was 800, 800, 300, and an output layer with 9, respectively. The activation function used in each hidden layer was the rectified linear unit (ReLU), ReLU, hyperbolic tangent (tanh), and linear function, respectively. The number of layers, number of neurons, and activation functions at each layer were optimized through trial and error based on the mean absolute (MAE) and root mean squared error (RMSE) of the verification set following training. The model was compiled using the Adaptive Moment Estimation (adam) optimization algorithm and the Huber loss function, as defined in Eq. (15),

$$L_\delta(y, \hat{y}) = \begin{cases} \frac{1}{2}(y - \hat{y})^2 & \text{if } |y - \hat{y}| \leq \delta \\ \delta(|y - \hat{y}| - \frac{1}{2}\delta) & \text{otherwise} \end{cases}, \tag{15}$$

where $y$ represents the true value, $\hat{y}$ represents the predicted value, and $\delta$ is a hyperparameter that controls the threshold between the linear and quadratic regions of the loss function. The second machine learning algorithm was implemented using the random forest regressor (RFR) class from the sklearn (or Scikit-learn) library with 50 decision trees (or estimators).

The input features were organized such that each column of the feature matrix (**X**) represented a feature and each row of the matrix represented a training structure. One row of the feature matrix included the atomic positions, their corresponding one-hot encoded atomic types, and the three lattice vectors ($\vec{a}$, $\vec{b}$, and $\vec{c}$). The atomic types were one-hot encoded based on the unique set of chemical symbols present in the dataset. Each atomic type was transformed into a binary vector where each element corresponds to the presence or absence of a particular chemical symbol (i.e., (1, 0, 0) for Mo, (0, 1, 0) for Re and (0, 0, 1) for W). These one-hot encoded vectors were then concatenated with the respective atomic positions, resulting in a comprehensive feature set that includes spatial and type-specific information for each atom. The features were flattened, creating a feature matrix with 1,014 rows and 297 columns. The 297 columns accounted for 3 Cartesian coordinates and 3 one-hot encoded atomic type values multiplied by 48 to account for each atom and 9 elements representing the lattice vectors (48·3 + 48·3 + 9). Both models showed very little sensitivity to re-ordering the input features. Please note, additional features were trialed and ruled out due to lack of any significant improvement in the model's performance (less than 2 GPa improvement in the MAE or RMSE). This included nearest neighbor information such as atomic type, bond length, and bond angle and principle component analysis for the atomic positions. Any "calculated" (such as energy or stress) features were purposefully left out as the key motivation to training the model was to circumvent any need to perform a calculation on new structures. In this way, by simply generating a POSCAR file, the models will deliver an estimated elastic constants prediction. The target value matrix (**y**) contained non-zero (9 out of the 21) elastic constants, as calculated by ElasTool, for each training structure (1,014 rows and 9 columns). Both **X** and **y** were standardized using the StandardScaler preprocessing technique which was saved alongside the trained model.

The k-fold cross-validation method was used to assess the performance of the two machine learning models during the fitting process. The k-fold technique was implemented with k = 10 splits for the k.Seq and RFR models. Cross-validation is a widely used technique in machine learning that helps evaluate the model's generalization ability and mitigate potential biases. The k-fold cross-validation approach involved partitioning the available dataset into k equally sized subsets (or folds). The model was then trained and evaluated k times, with each fold serving as the validation set once and the remaining nine folds used for training. By repeatedly cycling through different combinations of training and validation sets, it was possible to obtain a more robust and reliable estimate of the model's performance. This approach allowed for an assessment of the model's ability to generalize to unseen data and provided a more comprehensive evaluation of its performance. To avoid over-fitting when training the k.Seq model, an early stopping criteria was introduced based on the progression of the validation loss value with a patience setting of five epochs and a maximum number of 100 epochs allowed per fold. If the validation loss value did not continue to decrease after five epochs the model weights reverted to those which resulted in the best validation loss value. The patience setting was optimized through trial and error based on the RMSE value of the verification sets. Given a training set sized at 1,014 points, there were approximately 101 validation structures and approximately 913 training structures at each fold.

## 3. Results and discussion

### 3.1. Material geometry and composition

(MC)$^2$ predicted a majority single BCC phase for all four alloys, referred to as a, b, c, and d, at LT and HT which was confirmed experimentally. The lattice constant values were in good agreement with the experimentally measured values as the relative error was below 1% for each alloy. (MC)$^2$ showed a thermodynamic preference for the formation of (W,Re) precipitated phases at LT and HT. Experimentally, the alloy samples did show evidence of precipitates after high temperature exposure. This highlights the power of (MC)$^2$ to enhance discovery and characterization of hard-to-measure phases due to their low volume fractions. Despite having low molar fractions, precipitated phases can have profound impacts on a structure's mechanical properties, especially in HT environments. [22,59–61] A discussion on the precipitates properties is included in the machine learning section. The (MC)$^2$ phases are listed in Table 2. The BCC structures are well supported considering the (Re,W) [62], (Mo,W) [63], and (Mo,Re) [64] binary phase diagrams and the (Mo,Re,W) ternary phase diagram at HT [65,66]. The one exception is phase b', which could evolve into the brittle intermetallic sigma phase [62,67] at temperatures near LT. However, coexistence with the dominant (Mo,W) matrix could stabilize a BCC matrix for phase b'. It is important to note that the precipitated (W,Re) phases should provide an enhancement to mechanical performance at HT. The majority phase predictions are in excellent agreement with the ternary diagram at HT which shows a (Mo,W) BCC phase where Mo, W, and Re freely occupy sites of the BCC lattice with no specific order.

(MC)$^2$ indicated a thermodynamic preference for the formation of precipitated phases in an equilibrated solid solution. Alloy a, which had the lowest amounts of W and Re, formed a precipitate with lowest Re content. When the W and Re content was increased in alloy b there was a thermodynamic drive for the formation of a richer Re precipitate, $WRe_{25}$. This would indicate that, given enough W and Re within the matrix, there is a tendency for W and Re to aggregate. However, in alloy c the W content is doubled in favor of a near-equal blend of Mo and W while the Re content increased by only 3 at.%. Phase c forms the (Mo,W) majority phase, but with higher W content which created a mixed BCC lattice with higher Re content. In this case, we see a thermodynamic preference where Mo aggregates with W and Re aggregates with W. For alloy d, there is also the formation of the (Mo,W) system with Re mixed in and a small amount of a $WRe_4$ precipitate. The high resolution thermodynamic phase stability findings indicated a balance between Mo and W in order to promote (Mo,W) and (W,Re) phases. Molybdenum played an important role in stabilizing the (Mo,W) BCC matrix with higher W content which promoted further Re diffusion into the (Mo,W) matrix. Phase b' was the richest in Re and it is interesting to note that alloy b had the highest Re:W content ratio. In fact, alloy a and d had the lowest Re:W ratios and lowest Re content in the WRe precipitates. The phase predictions are in good agreement with experimentally generated phase diagrams and offer interesting insight into the phase stability dynamics between (Mo,W) and (W,Re) as both Mo and Re showed a strong thermodynamic preference to aggregate with W. By considering W the prime element, it becomes a balancing act of Mo:Re content to promote phase stability of either the (Mo,W) or (W,Re) system. This level of fine-detailed analysis is crucial to guide alloy development for tailor-made high temperature applications.

The elastic properties for the majority phases predicted by (MC)$^2$ at LT and HT are compiled in Table 3a. For accuracy validation, Table 3b provides experimental references. The mechanical properties are consistent with experimental data. Notably, alloy c exhibited the least degradation between low and high temperatures. This is attributed to its (Mo,W) matrix with the highest W and Re content among the blends. This suggests that alloy c could be the preferred blend for high temperature applications. Melting temperature and micro-hardness predictions confirm these alloys are suitable for abrasive high-temperature applications. Fig. 2a shows the LT elastic moduli for each alloy. The increase in bulk modulus (B), shear modulus (G), and Young's modulus (Y) with increasing W content indicates that the Mo-Re-W system became stiffer and more resistant to deformation. The higher B suggests greater resistance to uniform compression, reflecting stronger inter-atomic bonding. The increase in G implies enhanced resistance to shape changes under shear stress, indicating improved rigidity. Similarly, a higher Y denotes greater stiffness in response to tensile stress, signifying that the material was more resistant to stretching. Fig. 2b shows that alloys c and d retained 90% of their LT B, G, and Y values when at HT, indicating exceptional high-temperature stability and mechanical robustness. This retention of elastic properties suggests that these alloys have strong resistance to thermal softening and maintain their stiffness and rigidity even at extreme temperatures. Consequently, alloy c is likely to perform reliably in high-temperature applications, such as aerospace or power generation, where maintaining mechanical integrity under thermal stress is critical [1,6,7,41]. Fig. 2c shows a decrease in $\Theta_D$ vs. $k_{min}$ with increasing W

**Table 2**
A list of the computationally predicted phases for the alloys at LT and HT. The primed phases represent the precipitated phases that were predicted to from at LT and HT. Alloy composition is provided in atomic percent (at.%) and molar fraction (m.f.) values are included in parenthesis for co-existing phases.

| Ref. | Crystal Structure | | Alloy Comp. (at%) | |
|---|---|---|---|---|
| | P1(m.f.) | P2(m.f.) | P1 | P2 |
| a, a' | BCC(96) | BCC(4) | $MoRe_6W_8$ | $WRe_2$ |
| b, b' | BCC(96) | BCC(4) | $MoRe_{12}W_{19}$ | $WRe_{25}$ |
| c, c' | BCC(94) | BCC(6) | $MoRe_{17}W_{35}$ | $WRe_{10}$ |
| d, d' | BCC(94) | BCC(6) | $MoRe_8W_{56}$ | $WRe_4$ |

**Table 3**

(a) The computed elastic properties are in units of GPa and unitless, respectively, up to the Kleinman parameter ($\xi$). Following $\xi$, the units are K, K, and W/m.K, respectively. (b) A compilation of the elastic properties for molybdenum, tungsten, rhenium, a molybdenum-rhenium (MoRe$_{48}$) alloy, and tungsten-rhenium (WRe$_{25}$) alloy using the same units as sub-table (a).

| (a) Alloy | B | G | Y | C" | H | G/B | $\nu$ | $\xi$ | $\Theta_D$ | $T_m$ | $k_{min}$ |
|---|---|---|---|---|---|---|---|---|---|---|---|
| a | 279 | 128 | 333 | 75.7 | 17.4 | 0.460 | 0.301 | 0.600 | 437 | 3309 | 1.27 |
| a$_{HT}$ | 240 | 115 | 297 | 58.3 | 15.3 | 0.478 | 0.294 | 0.581 | 414 | 2945 | 1.19 |
| b | 286 | 132 | 342 | 74.7 | 17.9 | 0.459 | 0.301 | 0.638 | 417 | 3305 | 1.21 |
| b$_{HT}$ | 232 | 113 | 291 | 50.9 | 14.9 | 0.485 | 0.291 | 0.607 | 385 | 2806 | 1.10 |
| c | 296 | 137 | 355 | 72.3 | 18.5 | 0.462 | 0.300 | 0.668 | 398 | 3330 | 1.15 |
| c$_{HT}$ | 263 | 125 | 323 | 58.5 | 16.7 | 0.476 | 0.295 | 0.651 | 380 | 3031 | 1.09 |
| d | 299 | 143 | 370 | 66.8 | 19.1 | 0.478 | 0.294 | 0.641 | 392 | 3430 | 1.12 |
| d$_{HT}$ | 259 | 128 | 330 | 50.2 | 16.7 | 0.495 | 0.287 | 0.623 | 371 | 3051 | 1.05 |
| (b) Alloy | | | | | | | | | | | |
| Mo [68] | 260 | 126 | 329 | | 16.9 | 0.485 | 0.310 | | | | |
| W [69] | 310 | 161 | 411 | | 20.4 | 0.519 | 0.280 | | | | |
| Re | 370 | 178 | 463 | | 23.8 | 0.481 | 0.300 | | | | |
| MoRe [70] | | 132 | 365 | | | | 0.285 | | | | |
| WRe [70] | | 159 | 430 | | | | 0.290 | | | | |

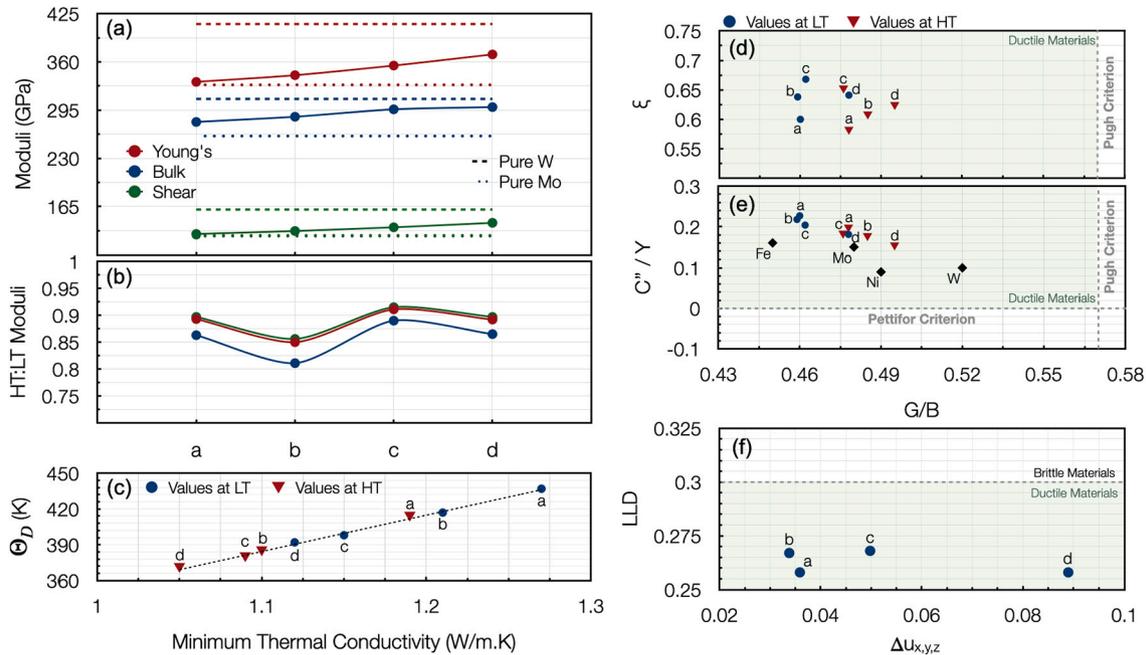

**Fig. 2.** (a) Elastic moduli at LT for the studied alloys, with pure W (dashed line) and pure Mo (dotted line) for comparison. (b) Ratio of HT to LT elastic moduli for B, G, and Y, showing that all alloys maintain between 80 and 90% of their LT properties at HT, indicating high-temperature resilience. (c) Debye temperature ($\Theta_D$) vs. minimum thermal conductivity ($k_{min}$) for LT and HT. Note the clear correlation with these properties and Mo content. (d) Kleinman parameter ($\xi$) vs. Pugh ratio (G/B). (e) Normalized Cauchy pressure (C") by Young's modulus (Y) vs. G/B, with the ductile-to-brittle boundary at 0.57 and Pettifor's limit. Pure metals included for comparison from Ref. [71]. (f) Local lattice distortion vs. average atomic displacement, with the brittle-to-ductile transition value of 0.3 marked by a horizontal dashed line.

content and a corresponding increase with Mo content. This provides insights into the material's thermal and structural characteristics. Lower $\Theta_D$ with higher W content suggests W atoms introduce softer bonds and lower vibrational frequencies, leading to reduced phonon velocities and hence lower thermal conductivity. Conversely, higher $\Theta_D$ with increased Mo content indicates Mo atoms enhance bond stiffness, resulting in higher vibrational frequencies and increased phonon velocities. This leads to improved thermal conductivity, reflecting stronger interatomic interactions and greater material stiffness. Therefore, the balance between W and Mo content can be strategically adjusted to tailor the alloy's thermal and mechanical performance for specific high-temperature applications, such as aerospace components (e.g., leading-edge for high speed vehicles) [41,72,73] or nuclear reactor materials [74].

Figs. 2d, e, and f show that all considered alloys are ductile. The magnitude of $\xi$ indicates a preference for deformation through bond bending over stretching, with Pugh ratios within the ductility region. Alloy a's low $\xi$ value supports the finding that Mo increases the rigidity of the Mo-Re-W lattice, enhancing thermal properties. Comparing LT and HT positions in Figs. 2d and e, all systems became more brittle at HT. Notably, alloy c showed the least degradation in ductility and stiffness from LT to HT. The alloys' ductility compared to pure Mo or W (Fig. 2e) demonstrates that even small Re additions enhance ductility. Alloy d's high Y values and increased lattice movement (Fig. 2f) suggest good mechanical integrity under stress but potential challenges in maintaining structural stability under thermal cycling or long-term exposure to high temperatures. Increased displacement could lead to

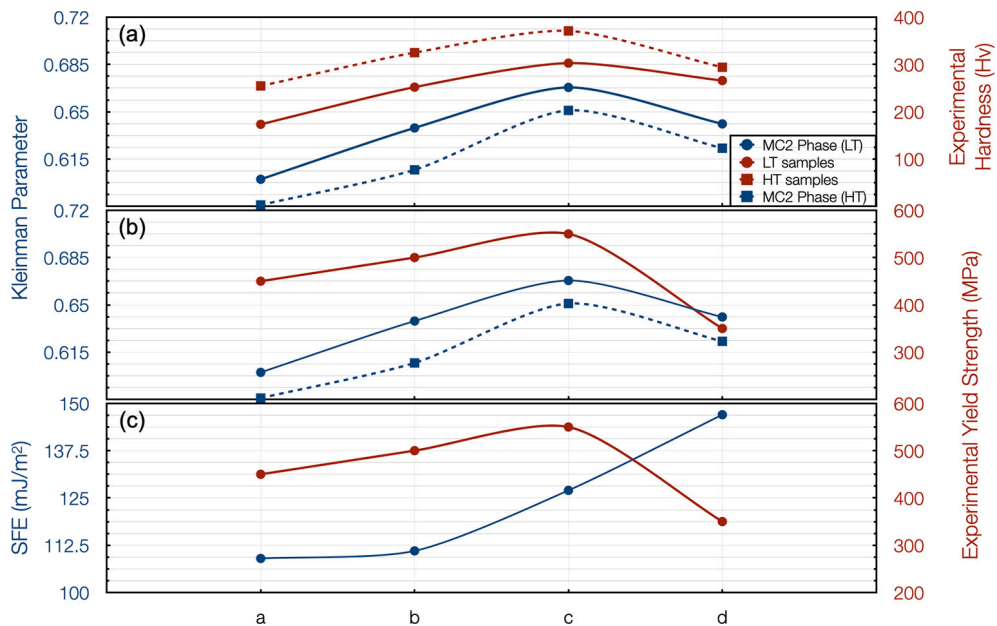

**Fig. 3.** Variations in the Kleinman parameter ($\xi$), experimental hardness, yield strength, and stacking fault energy (SFE) between the different alloy compositions. (a) $\xi$ is plotted for the different alloys (a, b, c, d) at LT and HT. The right y-axis represents experimental hardness (Hv). (b) $\xi$ is plotted alongside the experimental yield strength (MPa) for each alloy. (c) The stacking fault energy (SFE) plotted against yield strength.

higher internal stresses and potential defects, affecting overall durability and performance.

Experimental observations support the phenomenon of Re increasing a W alloy's ductility, as seen in the works of Lai et al. [29] and Kishore et al. [30], who reported smaller W grain sizes and improved ductility with increased Re content. Romaner et al. [75] discovered that alloying W with Re transformed the symmetric screw dislocation core into an asymmetric configuration. This transformation shifted the slip system from 110 to 112, providing six additional slip planes and reducing the Peierls stress from 2.49 GPa for pure W to 1.84 GPa for the W-25 at.% Re alloy, resulting in a more ordered grain structure. Chen et al. [28] observed smaller W grain sizes and improved ductility by adding up to 10 wt.% Mo to the W-Ni-Fe WHA. Similarly, Zhang et al. [76] examined the effects of Mo in the $Nb_{30}Ta_{30}Mo_x(Ti_2Ni)_{40-x}$ HEA (x = 0, 10, 20) and concluded that grain size was reduced with the addition of Mo, yielding a mechanically hardened alloy with reduced plasticity. A 2016 DFT study [26] concluded that Re added to pure W or Mo enhanced grain boundary cohesion, reducing the likelihood of intergranular fracturing. The phase stability analysis conducted within this work showed that Mo introduces a strong thermodynamic preference for the formation of a (Mo,W) lattice, likely contributing to reduced W grain size. Findings here are consistent with the observation that increased Re content enhances ductility while increased Mo content reduces plasticity (Figs. 2c, d, e, f). Furthermore, the phase, thermal, and elastic findings here are in good agreement with a recent DFT study on the impact of doping brittle materials with transition metals (such as Mo, Re, W) [38], which concluded that these additions enhance mechanical properties, reduce hardness, and improve thermodynamic phase stability. By integrating these experimental and computational insights, there is a clear trend: the addition of Re and Mo to W alloys not only modifies grain structure but also significantly impacts ductility and mechanical performance, making these alloys highly suitable for extreme high temperature applications.

### 3.2. Mechanical predictions

Fig. 3 presents an analysis of the mechanical properties and structural parameters of the four different compositions of the Mo-Re-W ternary alloy system. In Fig. 3a and b, $\xi$ is compared against the experimental hardness and yield strength. It can be seen that $\xi$ increases as Mo:W content nears 1:1, up to alloy c, and then decreases at alloy d where W becomes the majority constituent and Re content decreases from 16 at.% to 8 at.%. This suggests that a combination of increased W and Re enhances the material's resistance to stretching, favoring bending which is optimal in alloy c. This balance in deformation behavior is critical for applications requiring both flexibility and strength. Experimental hardness and yield strength increased with W and Re content, as well, peaking at alloy c and decreasing at alloy d. This trend highlights the role and critical balance of Mo, W, and Re in enhancing the stiffness and mechanical strength of the alloy, which is vital for structural applications that demand high durability and load-bearing capacity at elevated temperatures. The presence of W and Re contributed to solid solution strengthening and increased the overall stiffness of the alloy, making it suitable for high-performance aerospace components where both high strength and resistance to thermal softening are crucial [41,77]. As seen in Fig. 3c, the SFE increased with W content all the way to alloy d. Higher SFE values facilitate dislocation movement, which can enhance ductility but may also reduce the material's resistance to plastic deformation. The higher SFE indicated increased dislocation mobility in alloy c and d, which is beneficial for applications requiring materials that can accommodate thermal expansion and contraction without cracking, such as aerospace applications [24,41] and nuclear reactor components [77]. Alloys a and b had higher amounts of Mo and displayed lower SFE. This indicates that Mo increases the alloy's resistance to dislocation motion and enhances strength through solid solution strengthening. However, excessive Mo, as seen in alloy a, can reduce overall stiffness, affecting hardness and yield strength negatively. Molybdenum's contribution to lowering SFE is advantageous for applications requiring high strength and resistance to creep, such as power generation turbines [74,77]. Rhenium, present in smaller amounts, provided additional solid solution strengthening, enhancing both hardness and yield strength. Rhenium's influence was more pronounced in alloy c which had balanced W and Mo contents, contributing to the overall mechanical robustness of the alloy. Considering insights from the phase stability discussion, alloy c's better performance could be due to the fact that it developed a (Mo,W) matrix with the highest amount of both W and Re relative to alloys a and b.

**Table 4**

The percent difference between predicted and actual values of the 33 unseen verification structures obtained from the average root mean squared error (RMSE) for each elastic constant. The elastic moduli, Cauchy pressure, and Debye temperature are included to give a sense of error in some of the elastic and thermal properties that were obtained from the predicted elastic constants. The units for the elastic moduli are GPa and the Debye temperature is units K. The standard deviation in units GPa for each elastic constant across the 33 verification structures is included to demonstrate the variability of the verification data set.

|  | % Difference | | | | | |
|---|---|---|---|---|---|---|
|  | k.Seq | RFR | $\sigma$ |  | k.Seq | RFR |
| $C_{11}$ | 1.56 | 1.72 | 11.1 | B | 0.84 | 0.71 |
| $C_{12}$ | 2.81 | 1.59 | 20.6 | G | 1.89 | 1.63 |
| $C_{13}$ | 1.74 | 1.27 | 20.7 | Y | 1.66 | 1.45 |
| $C_{22}$ | 1.81 | 1.46 | 11.1 | C" | 6.11 | 3.95 |
| $C_{23}$ | 3.26 | 3.61 | 20.5 | $\Theta_D$ | 0.90 | 0.77 |
| $C_{33}$ | 1.76 | 1.79 | 14.7 |  |  |  |
| $C_{44}$ | 2.26 | 1.56 | 15.2 |  |  |  |
| $C_{55}$ | 1.34 | 1.28 | 15.2 |  |  |  |
| $C_{66}$ | 3.56 | 3.84 | 14.7 |  |  |  |

*3.3. Machine learning*

The average percent difference between the model predicted and actual value for each of the 9 elastic constants was calculated using the RMSE for the 33 verification structures and is provided in Table 4 for both models. The standard deviation of the actual values has been provided, as well, to provide insight into the variability of elastic constants in the verification set as obtained through spin-polarized DFT calculations and ElasTool. The coefficient of determination (commonly referred to as $R^2$) was not considered as a metric of performance due to its sensitivity to the low variability in the input features and the target values across the training set. It is important to stress that, while it is interesting to see how each model performed side-by-side, the information provided in Table 4 should not be used to suggest that one model performed better than the other. When it comes to model complexity, the RFR and k.Seq models have different levels of complexity. The RFR model is a tree-based ensemble model, while the k.Seq model is a deep neural network. It is highly likely that comparing the two models based solely on MAE and RMSE would not capture the full complexity of the models. When considering the interpretability of each model, RFR models are often easier to interpret due to their decision tree structure. On the other hand, deep neural networks can be very complex, making it challenging to understand how they arrive at their predictions. The interpretability aspect is not captured solely from the RMSE. Rather, this exercise demonstrates that both models, when trained and verified on identical data sets, are capable of generating realistic elastic constant predictions. The RFR model is a much more user-friendly option as it has far less parameters to tune (number of estimators). Whereas with k.Seq, the user is free to alter the number of nodes, number of layers, activation functions, loss functions, optimizer options, and so on. An adjustment to any one of the aforementioned features will vary the k.Seq model's performance.

Confident in the models' performance, VASP and ElasTool were replaced by an "elastic constants calculator" and elastic constants were generated for 615 never-before-seen BCC SCRAPs. To understand the relationship between each metallic constituent and the elastic and thermal properties at a well-resolved level, the 615 new structures were generated with an iterative sweep through the Mo-Re-W compositional domain from 10 at.% to 90 at.% with a step size of 2 at.% for each constituent. For this exercise the RFR model was used as the "elastic constants calculator". Together with the original 1,014 training structures, a total of 1,629 elastic constants were compiled. With a larger database spanning both a wide area of the atomic concentrations and many different atomic configurations, a generalized discussion on how each atomic species impacts the elastic properties of the Mo-Re-W system was possible. Fig. 4 provides a set of heat maps which was generated using the compiled library of 1,629 elastic constants. These maps illustrate the dependence of various elastic and thermal properties on the alloy constituents in the Mo-Re-W ternary alloy system. Each column corresponds to a different property (from left to right): Y, G/B, $\xi$, and $\Theta_D$. Each row corresponds to either W (top) or Re (bottom) content vs. Mo content. The positions of alloys a, b, c, and d are marked on the maps. Maps like these provide a powerful tool for designing tailored alloy blends, particularly for high-temperature applications. By visualizing the dependence of key mechanical and thermal properties on the concentrations of Mo, W, and Re, these maps facilitate the identification of preferred alloy compositions for specific use cases. For example, hypersonic aerospace vehicles operate under extreme temperatures and stresses, requiring materials that maintain mechanical strength and thermal stability. Using the heat maps, alongside density and cost considerations, an optimal composition like alloy c (Mo 45%, Re 15%, W 40%) can be identified. This alloy offers a balance of high stiffness, adequate ductility, and excellent thermal stability. These properties ensure that components can withstand high operational temperatures without losing structural integrity, improving performance and lifespan. The balanced Mo and W content enhances stiffness, strength, and phase stability, while Re improves ductility and resistance to thermal softening, making this composition particularly well-suited for the demanding conditions of hypersonic flight [41]. Additionally, this work discussed the balance between Mo and Re content and the thermodynamic preference to form either a (Mo,W) matrix or a (Mo,W) matrix with (W,Re) precipitates. Due to the (W,Re) system's superior high-temperature qualities, and the profound impact precipitated phases can have in high-temperature environments [22,60,59,61], blends approaching equiatomic concentrations should receive further enhancements for high temperature applications. This is achieved through the well-balanced thermo-mechanical properties of the primary (Mo,W) matrix and a further development of the (W,Re) precipitate system.

**4. Conclusion**

This work introduced a computational routine for exploring next-generation alloys using $(MC)^2$ and ElasTool to output meaningful parameters for high-temperature applications. Validated against experimental measurements and phase diagram data, the method provided meaningful insights and predictions on phase stability, elastic response, and thermal properties at room temperature and 1600 °C for a set of four ternary Mo-Re-W alloys. The $Mo_{45}Re_{16}W_{39}$ composition exhibited the best performance due to its balanced Mo and W blend promoting a BCC matrix and (W,Re) precipitates which bolstered high-temperature performance.

To circumvent the need to perform costly DFT calculations, 1,014 Mo-Re-W structures were generated and used along with their elastic constants to train a sequential deep learning (k.Seq) and random forest regressor (RFR) model. Both models showed accurate predictions and the RFR model was used to generate 615 new elastic constants spanning the Mo-Re-W compositional domain with a fidelity of 2 at.% per constituent.

A total of 1,629 elastic constants were compiled and displayed through heat maps that illustrated the relationship between the atomic constituents and the alloy's properties. These maps, coupled with $(MC)^2$ phase stability insights, demonstrated the power of machine learning to enable a holistic and statistically robust visualization of how each element impacts mechanical performance enabling quick identification of promising alloy blends for specific high temperature applications. Notably, near-equiatomic (Mo,W) blends were highlighted for their strong

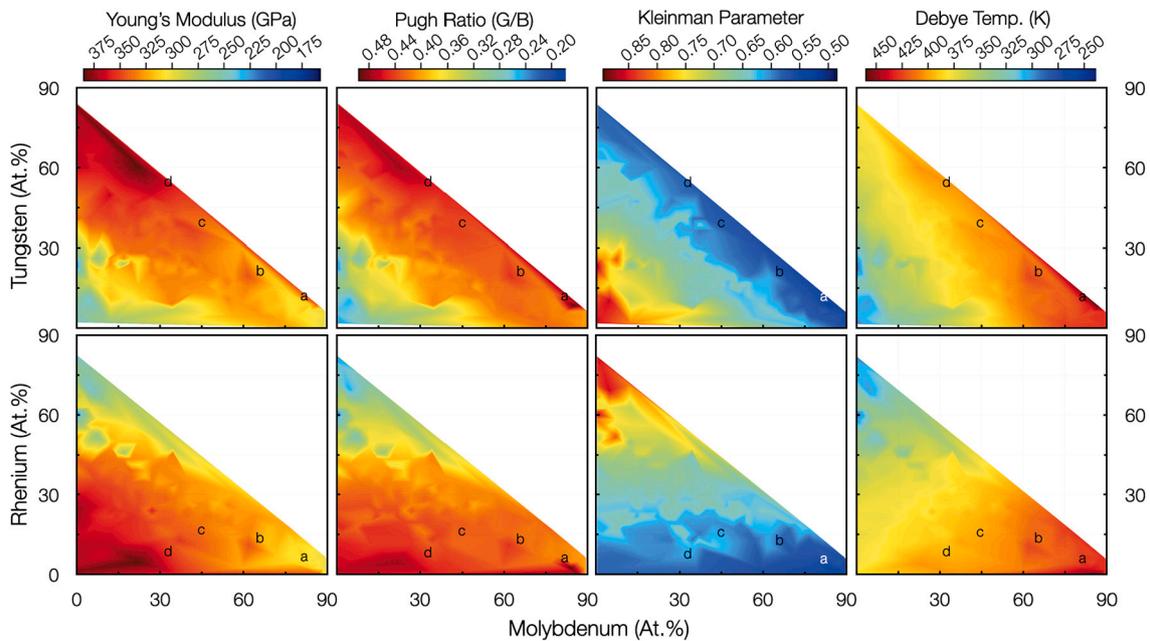

**Fig. 4.** Heat maps which correlate the atomic percents (At.%) of tungsten, top panel, and rhenium, bottom panel, against molybdenum for the Young's modulus, Pugh ratio, Kleinman parameter, and Debye temperature. The location of the alloys examined in this work are labeled a-d. These heat maps were generated using 1,629 structural predictions made by the random forest regressor model.

mechanical properties and excellent phase stability at low and high temperatures.

This work hopes to provide the materials science community with efficient computational methods for predicting and analyzing promising high-temperature alloys, emphasizing the crucial role of machine learning in understanding and optimizing alloy compositions for tailor-made needs across various high-temperature applications.

**CRediT authorship contribution statement**

**Tyler D. Doležal:** Writing – review & editing, Writing – original draft, Visualization, Validation, Software, Resources, Methodology, Investigation, Formal analysis, Data curation, Conceptualization. **Nick A. Valverde:** Validation, Methodology. **Jodie A. Yuwono:** Writing – review & editing, Methodology. **Ryan A. Kemnitz:** Supervision, Project administration.

**Declaration of competing interest**

We declare that we have no conflicts of interest regarding the submission of this manuscript to Acta Materialia. We affirm that the research presented in this manuscript was conducted in an impartial manner, and there are no financial, personal, or professional relationships that could influence the integrity or objectivity of our work.

**Data availability**

The (MC)$^2$ code used for this work is available at https://github.com/SaminGroup/Dolezal-MC2. The data, scripts, training set and models developed as a result of this work will be made available at https://github.com/tylerdolezal.

**Acknowledgements**

This work was fully supported by computational resource allocations provided by the Department of Defense high performance computing through AFRL's HPC Mustang. This research was supported by the Air Force Research Laboratory's Materials and Manufacturing Directorate.